\documentclass[12pt,epsf]{elsart}
\usepackage{amsmath}

\begin{document}

\begin{frontmatter}

\title{Annealing of radiation induced defects in silicon
in a simplified phenomenological model}

\author[iftm]{S. Lazanu} and
\author[univ]{I. Lazanu}

\address[iftm]{National Institute for Materials Physics,
P.O.Box MG-7, Bucharest-Magurele, Romania, 
electronic address: lazanu@alpha1.infim.ro}
\address[univ]{University of Bucharest, Faculty of Physics,
P.O.Box  MG-11, Bucharest-Magurele, Romania,
electronic address: ilaz@scut.fizica.unibuc.ro}

\begin{abstract}

The concentration of primary radiation induced defects has been 
previously estimated considering both the explicit mechanisms of the 
primary interaction between the incoming particle and the nuclei of the 
semiconductor lattice, and the recoil energy partition between ionisation 
and displacements, in the frame of the Lindhard theory. 
The primary displacement defects are vacancies and interstitials, that 
are essentially unstable in silicon. They interact via migration, 
recombination, annihilation or produce other defects.
In the present work, the time evolution of the concentration of defects 
induced by pions in medium and high resistivity silicon for detectors is 
modelled, after irradiation. In some approximations, 
the differential equations representing the time evolution processes 
could be decoupled. The theoretical equations so obtained are solved 
analytically in some particular cases, with one free parameter, for a 
wide range of particle fluences and/or for a wide energy range of the 
incident particles, for different temperatures; the corresponding 
stationary solutions are also presented.  



61.70.At: Defects formation and annealing processes
\medskip
\end{abstract}

\end{frontmatter}

\section{Introduction}

A point defect in a crystal is an entity that causes an interruption in 
the lattice periodicity. In this paper, the terminology and definitions 
in agreement with M. Lannoo and J. Bourgoin \cite{1} are used in relation to 
defects.

Vacancies and interstitials are produced in materials exposed to 
radiation in equal quantities; they are the primary radiation defects, 
being produced either by the incoming particle, or as a consequence of 
the subsequent collisions produced by the primary recoil

In previous papers \cite{2,3} we calculated the concentration of 
radiation induced primary defects (CPD) in semiconductors exposed to 
hadron irradiation. Due to the important weight of annealing processes, 
as well as to their very short time scale, CPD is not a measurable 
physical quantity.

In silicon, vacancies and interstitials are essentially unstable and 
interact via migration, recombination, annihilation or produce other 
defects. The system evolves toward natural equilibrium.

The problem of the annealing of radiation induced defects in 
semiconductor materials is not new. Several models, empirical or more 
theoretic, have been previously proposed to explain these phenomena. 
Damask and Dienes developed, in the 1960s, a theoretical model that considers, as distinct cases, 
the kinetics of vacancy-interstitial annihilation with interstitial 
trapping at impurities \cite{4}, of vacancy-interstitial annihilation 
with interstitial trapping at impurities \cite{5}, and of 
vacancy-interstitial annihilation with diinterstitial formation \cite{6}.
In the first two cases, analytical solutions have been obtained by the 
authors, while in the last case only numerical solutions are possible. 
The escape of interstitials and vacancies from clusters and their 
subsequent interactions have been modelled by Tsveybak et al. \cite{7}. 
Phenomenological approaches of the annealing could be found in Moll et 
al. \cite{8} and Bates and co-workers \cite{9} and references cited therein. 

Most of the old studies were dedicated to electronic devices, 
made from low resistivity (in the $\Omega cm$ range) silicon. Recently, the 
behaviour of the silicon material with medium and high resistivity (tens 
$\Omega cm$ - $k\Omega cm$) in intense hadron fields started to present very much 
interest, these materials representing a major option for the detectors 
at the new generation of particle colliders and for space applications.

In the present paper, the time evolution of the concentration of primary 
defects induced by pion irradiation in silicon is studied in the frame 
of a simplified phenomenological model based on direct interactions 
between the primary induced defects and the impurities present in the 
material.

\section{Equations for the kinetics of radiation induced defects:
general formulation}

The starting hypothesis of the physical model is that the 
irradiation process is unique at the determined particle fluence, and 
that the time of irradiation is shorter that all characteristic times 
of annealing.

In the frame of the model, equal concentrations of vacancies and 
interstitials are supposed to be produced by irradiation, in much 
greater concentrations than the corresponding thermal equilibrium values, 
characteristic to each temperature. Both the pre-existing defects and 
those produced by irradiation, as well as the impurities, are assumed to 
be randomly distributed in the solid. An important part of the vacancies 
and interstitials annihilate. The sample contains certain concentrations 
of impurities which can trap interstitials and vacancies respectively, 
and form stable defects.

In the present paper, vacancy-interstitial annihilation, 
interstitial migration to sinks, vacancy and interstitial impurity 
complex formation as well as divacancy formation are considered. The 
sample could contain more impurities that trap vacancies or 
interstitials, and in this case all processes are to be taken into 
account.

The following notations are used: $V$ - monovacancy concentration; 
$I$ - free interstitial concentration, $J_1$ - total impurity "1" 
concentration (impurity "1" traps interstitials and forms the 
complex $C_1$); $C_1$ - interstitial-impurity concentration: one 
interstitial 
trapped for one complex formed; $J_2$ - total impurity "2" concentration 
(impurity "2" traps vacancies and forms the complex $C_2$); $C_2$ - 
vacancy-impurity concentration: one vacancy trapped for one complex 
formed; $V_2$ - divacancy concentration. All concentrations are expressed 
as atomic fractions.

The chemical scheme for reaction rates has been used, and the 
differential equations corresponding to the above picture are:

\begin{equation}							
\frac{dV}{dt}=-K_1VI-K_5V\left( J_{20}-C_2\right) +K_6C_2-K_7V^2+K_8V_2
\end{equation}

\begin{equation}                                                        
\frac{dI}{dt}=-K_1VI-K_2I-K_3I\left( J_{10}-C_1\right) +K_4C_1
\end{equation}

\begin{equation}							
\frac{dC_1}{dt}=K_3I\left( J_{10}-C_1\right) -K_4C_1
\end{equation}

\begin{equation}							
\frac{dC_2}{dt}=K_5V\left( J_{20}-C_2\right) -K_6C_2
\end{equation}

\begin{equation}							
\frac{dV_2}{dt}=\frac 12K_7V^2-\frac 12K_8V_2
\end{equation}

The bimolecular recombination law of interstitials and vacancies is 
supposed to be a valid approximation for the present discussion, because 
at the concentrations of vacancies of interest, only a small fraction of 
defects anneal by correlated annihilation if their distribution is 
random (see the discussion in \cite{10}).

If $N$ is the total defect concentration, expressed as atomic fraction:

\begin{equation}
N=V+I+2V_2+C_1+C_2							
\end{equation}

then it satisfies the differential equation:

\begin{equation}
\frac{dN}{dt}=-2K_1VI						
\end{equation}

The initial conditions, at $t=0$, are: at the end of the irradiation, there 
are equal concentrations of interstitials and vacancies: $I_0=V_0$;  
the concentrations of impurities are $J_{10}$ and $J_{20}$ respectively,
there are no complexes in the sample: $C_{10}=C_{20}=0$, and no 
divacancies $V_{20}=0$.

The reaction constants $K_1$ and $K_3$ are determined by the diffusion 
coefficient for the interstitial atom to a substitutional trap, and 
therefore $K_1=K_3$:

\begin{equation}
K_1=30\nu \exp \left( -E_{i1}/k_BT\right)
\end{equation}

where $E_{i1}$ is the activation energy of interstitial migration and $\nu$ 
the vibrational frequency. The reaction constant $K_2$ (related to interstitial 
migration to sinks) is proportional to the sink concentration $\alpha$:

\begin{equation}
K_2=\alpha \nu \lambda ^2\exp \left( -E_{i1}/k_BT\right)                                                 
\end{equation}

with $\lambda$ the jump distance.

$K_4$ characterises the decomposition of a complex into an impurity and an 
interstitial, and is given by:

\begin{equation}
K_4=5\nu \exp \left( \frac{E_{i1}+B_1}{k_BT}\right)							
\end{equation}

$K_5$ and $K_6$ describe the formation and decomposition of vacancy-impurity 
complexes, while $K_7$  and $K_8$ the formation and decomposition of 
divacancies. and are given by:
			
\begin{equation}
K_5=30\nu \exp \left( -E_{i2}/k_BT\right) 			
\end{equation}

with $E_{i2}$  the activation energy for vacancy migration;

\begin{equation}
K_6=5\nu \exp \left( -\frac{E_{i2}+B_2}{k_BT}\right)                                                             
\end{equation}

\begin{equation}	
K_7=30\nu \exp \left( -E_{i2}/k_BT\right)                                                      
\end{equation}

\begin{equation}                                                            
K_8=5\nu \exp \left( -\frac{E_{i2}+B_3}{k_BT}\right) 
\end{equation}

where $B_1$ is the binding energy of $C_1$, $B_2$ the binding energy of
$C_2$ and $B_3$ is the 
corresponding binding energy of divacancies.

Due to the mathematical difficulties to solve analytically the complete 
differential equation system, some simplifications are necessary.

\section{Hypothesis, approximations and discussions}

The interstitials are much more mobile in silicon in respect to vacancies, 
and are characterised by an activation energy of migration a factor of 
two times smaller.

This difference between the activation energies $E_{i1}$ and $E_{i2}$ 
respectively justifies the introduction of two time scales, and the 
separate study of the processes involving interstitials and vacancies 
respectively. In the first stage, vacancy interstitial annihilation and 
interstitials migration to sinks are studied. The concentration of 
interstitials decays much rapidly than the concentration of vacancies. 
A cut-off condition for $I$ is imposed (a $p$ times decrease 
of interstitial concentration, $p$ being an adjustable 
parameter of the model). The vacancy concentration determined by this 
procedure is the initial value for the processes comprised in the second 
stage, and will be denoted by $V_{initial}$. The formation of 
complexes by vacancies is considered less important, and is neglected in 
the following discussion.

So, for the "first stage", after some simple manipulations, from 
(1) and (2), the equation:

\begin{equation}
\frac{dI}{dt}=1+\frac{K_2}{K_1V}
\end{equation}

has been obtained, with the solution:

\begin{equation}
I=V+\frac{K_2}{K_1}\ln \frac V{V_0}
\end{equation}

and:

\begin{equation}
t=\frac{\ln \left[ 1+\frac{K_1V\left( t\right) }{K_2\ln \left( V\left(
t\right) \right) }\right] }{K_2\ln \left( V\left( t\right) \right) }
\end{equation}

Imposing the cut-off condition for the concentration of interstitials, 
both $V_{initial}$ and the characteristic time could be found. As 
specified, $V_{initial}$ is used 
as initial vacancy concentration for the second step in the analysis, 
where the equations (4) and (5) are considered. 
$V_{initial}$ depends on $V_0$ and $p$, and it is temperature independent. 
This system of equations, 
expressing the kinetics of divacancy and vacancy-impurity formation, 
have no analytical solution. These processes are governed by the initial 
concentrations of vacancies and impurities.

If the impurity that traps vacancies is phosphorus, the limiting cases 
correspond to low initial doping concentration (high resistivity, 
uncompensated materials) and very high impurity concentration (low 
resistivity), respectively. In both cases the equations could be 
decoupled, and analytical solutions are possible. 

So, if the formation of vacancy-impurity complexes is not so important, 
the main process responsible for the decay of vacancy concentration is 
divacancy production. In this case, the time evolution of the vacancy 
concentration is described by the equation:

\begin{equation}
V\left( t\right) =\frac 1{4K_7}\left\{ -K_8+R\frac{\frac{K_8+4K_7V_i}%
R+\tanh \left( \frac{tR}4\right) }{1+\frac{K_8+4K_7V_i}R\tanh \left( \frac{tR%
}4\right) }\right\} 
\end{equation}

where:

\begin{equation}
R\equiv \sqrt{K_8\left( K_8+8K_7V_i\right) }
\end{equation}

while the increase of the divacancy concentration is given by:

\begin{equation}
V_2\left( t\right) =\frac{V_i-V\left( t\right) }2
\end{equation}

The stationary solution for $V(t)$ is given by:

\begin{equation}
\lim_{t\rightarrow \infty} V\left( t\right) =\frac1{4K_7}\left( R-K_8\right)
\end{equation}
	
For n-type high doped Si, the process described by eq. (4) is 
the most probable. If $J_0$ is the initial concentration of impurities, 
and the initial concentration of complexes is zero, than:

\begin{equation}
V\left( t\right) =\frac 1{2K_5}\left\{ -K_6+K_5\left( V_i-J_0\right) +R^{*}
\frac{\frac{K_6+K_5\left( V_i-J_0\right) }{R^{*}}+\tanh \left( \frac{tR^{*}}%
2\right) }{1+\frac{K_6+K_5\left( V_i-J_0\right) }{R^{*}}\tanh \left( \frac{%
tR^{*}}2\right) }\right\} 
\end{equation}

with:

\begin{equation}
R^{*}\equiv \sqrt{K_6^2+K_5^2\left( V_i-J_0\right) ^2+2K_5K_6\left(
V_i+J_0\right) }
\end{equation}

and with the stationary solution:

\begin{equation}
\lim_{t\rightarrow \infty} V\left( t\right) =\frac 1{2K_5}\left[ K_5\left( V_i-J_0\right)
+R^{*}-K_6\right] 
\end{equation}

and 

\begin{equation}
C=V_i-V
\end{equation}

\section{Results and physical interpretations}

In Figure 1, the absolute value of the CPD per unit fluence (CPD) 
induced by pions in silicon is presented as a 
function of the kinetic energy of the particle. This curve has been 
obtained in the hypothesis listed in references \cite{11,12}. In the 
kinetic energy range between 100 to 300 MeV, pion - nucleus interaction 
is resonant in all waves, the $\Delta_{33}$ resonance is produced and a 
pronounced maxima in the CPD is observed at approximately 150 MeV 
followed by a slight monotonic decrease at higher energies. Some local 
maxima are also presents at high energies, but with less importance. 

The process of partitioning the energy of the recoil nuclei (produced 
due the interaction of the incident particle with the nucleus, placed in 
its lattice site) in new interaction processes, between electrons 
(ionisation) and atomic motion (displacements) has been considered in 
the frame of the Lindhard theory.

The CPD multiplied by the fluence is the initial value of the 
concentration of vacancies and interstitials, and in the forthcoming 
discussion it is expressed, as specified, as atomic fraction. Without 
loss of generality, will shall consider that the primary defects are 
produced by 150 MeV kinetic energy pions.

The following values of the 
parameters have been used:  $E_{i1}$ = 0.4 eV, $E_{i2}$  = 0.8 eV,  
$B_1$=  0.2 eV, $B_2$  = 0.2 eV, $B_3$ = 0.4 eV, $\nu$ = 10$^{13}$ Hz,
$\lambda^2$  = 10$^{15}$ cm$^2$, $\alpha$  = 10$^{10}$ cm$^{-2}$, 
The value of the parameter$p$ is taken 100.

The time evolution of the concentration of vacancies, normalised to the 
concentration of vacancies created by irradiation is represented in 
Figure 2. Fluences in the range $10^{11} - 10^{15}$ pions/cm$^2$ have 
been considered. The weight of the annihilation process in respect to 
interstitial migration to sinks increases abruptly with the fluence. 
At low fluences, up to $10^{11}$ pions/cm$^2$, the annihilation has a 
low importance. With a good approximation, it could be considered that 
after 0.2 sec. after irradiation the vacancy concentration is saturated. 
As could be seen from the figure, observable effects in ${V}/{V_0}$ 
appear for fluences higher than $10^{12}$ pions/cm$^2$. Experimentally, 
for silicon particle detectors irradiated with hadrons, some 
modifications in the electrical characteristics have been observed in 
the fluence range $10^{12} - 10^{13}$ cm$^{-2}$ \cite{13,14}. 

All the curves correspond to room temperature (20 $^o$C) annealing.

The interval of time after which the concentration of interstitials 
decreases $p$ times, given by eq. (17) with $V(t) = V_{initial}$  is a 
characteristic time of the processes in the first stage, and depends on 
temperature and fluence. 

In Figure 3, this characteristic time is represented as a function of 
fluence, in the fluence range $10^{11} - 10^{15}$ pions/cm$^2$, for 
three temperatures of interest for high energy physics applications: 
-20, 0 and +20 $^o$C. It may be seen that, up to $10^{13}$ pions/cm$^2$, 
the characteristic time is independent on the fluence, and for higher 
fluences a decrease of the characteristic time is to be noted and a 
fluence dependence is obtained.

The initial concentration of vacancies for the second stage is 
$V_{initial}$  previously calculated. The second stage starts at a time 
equal to the characteristic time. 

Two limiting cases have been studied, low doping concentration 
(high resistivity uncompensated material), and low resistivity, 
phosphorus doped silicon. The formation and decomposition of divacancies 
is a good approximation for the first, while the second is related to 
$V-P$ complexes.

The time evolution of the concentration of vacancies when divacancy 
formation is considered is illustrated in Figure 4a. As underlined in 
Section 3, the time scale of this process is much longer in respect to 
the first stage. The origin of time in Figure 4a is the beginning of the 
second stage, this means a characteristic time after the start of the 
annealing.

The curves in this figure are normalised to the concentration of 
vacancies remained after the first stage. Values of the initial 
concentrations of vacancies that correspond to fluences in the same 
range as in the previous analysis have been considered. It could be 
observed that the time after which ${V(t)}/{V_{initial}}$ saturates 
is in the range $10^7 - 10^8$ sec., and increases with the irradiation 
fluence. All curves have been calculated also for 20 $^o$C temperature.

The time evolution of divacancy concentration is represented in Figure 
4b, also in a similar manner with the vacancy concentration in Figure 4a.

The case of high initial impurity (phosphorus) concentration, where 
divacancy formation is neglected in respect to complex formation, is 
presented in Figures 5a and 5b respectively.

Three initial concentrations for phosphorus in silicon (expressed as 
atomic fractions) have been considered: $10^{-5}$, $10^{-7}$, and $5\cdot 10^{-10}$, 
corresponding to 0.1 $\Omega cm$, 10 $\Omega cm$, and 1 $k\Omega cm$ 
resistivity respectively. The time evolution of the concentration of 
vacancies is represented in Figure 5a, while the corresponding one for 
the concentration of complexes in Figure 5b, both normalised to the 
concentration of vacancies remained after the first stage. An 
irradiation with $10^{13}$ pions/cm$^2$ has been considered. 

The origin of time in Figures 5a and 5b is displaced in respect with 
the origin of the annealing, in the same way as in Figs. 4a and 4b. 

The curves corresponding to $10^{11} - 10^{15}$ cm$^{-2}$ are 
superimposed on the same curve, with the exception of the tail at times 
longer than 10$^4$ sec. where the values corresponding to higher fluences 
are slightly higher. 

The time interval after which ${V(t)}/{V_{initial}}$ saturates is 
shorter than in the case of divacancy, and at room temperature is in 
the range $10^3 - 10^4$ sec. for the doping concentrations considered. 
It shortens with the increase of impurity concentration.

Some explicit considerations must be done about the formation of 
vacancy-impurity complexes. The mechanisms supposed above can be used 
both for boron and for phosphorus impurities. In this case, the 
corresponding processes are \cite{8}:

\begin{equation}
B_i+V \rightarrow \left[ B_i-V \right]
\end{equation}
                       
and respectively:

\begin{equation}
P_s+V \rightarrow \left[ P_s-V \right]
\end{equation}

While the complex formed by boron is unstable and self anneals bellow 
room temperature, the interaction between a $V$ and a $P_s$  leads to the 
formation of an $E$ centre which is stable in the same conditions \cite {7}.

Interactions between interstitial oxygen (another very studied impurity 
in the last time) and free vacancies is described as a higher order 
process (third order in the \cite{8} and fourth power in \cite{15}). If the 
process is stopped as a first order one, the time evolution of the 
concentrations is not different from the case of phosphorus, studied 
before. 

If two or more impurities that trap vacancies are considered as 
existing simultaneously in silicon, the system of coupled equations must 
be solved. Only numerical solutions for particular cases are possible.

If two or more impurities that trap vacancies are considered as existing 
simultaneously in silicon, the system of coupled equations must be 
solved. Only numerical solutions for particular cases are possible.

So, in the frame of this model, three characteristic times have been 
found for 20 $^o$C self annealing in silicon: one in the order of 
0.2 sec., corresponding to vacancy - interstitial annihilation and 
interstitial migration to sinks, one around 10$^4$ sec., related to 
vacancy-phosphorous formation and decomposition, and another one around 
10$^7$ sec., divacancy related. 

The present model could be applied to the annealing of primary radiation 
defects induced in Si by any other hadron, the difference coming from 
the value of the CPD.

\section{ Summary}

The time evolution of the primary concentration of defects induced by 
pions after irradiation process has been modelled.

In this model, vacancy-interstitial annihilation, interstitial migration 
to sinks, vacancy and interstitial impurity complex formation as well as 
divacancy formation have been considered. In the considered hypothesis, 
the time evolution of impurity concentrations has been decoupled into 
two steps, the first one involving vacancy-interstitial annihilation and 
interstitial migration to sinks, the second vacancy-complex and divacancy 
formation.

The equations corresponding to the first step have been solved 
analytically for a wide range of particle fluences and for a wide energy 
range of incident particles, and for three different temperatures: -20, 0 
and +20 $^o$C.

The approximations that permit to decouple the differential equations 
representing the time evolution processes in the second step have been 
studied, and the processes have been treated separately. The 
simultaneous consideration of more processes in this last step of the 
model is possible only numerically.

The predictions of the model are in general agreement with the 
experimental results, and suggest a possible connection with the self 
annealing mechanisms.

\section{Acknowledgements}

The authors are very grateful to Professor Gh. Ciobanu from the Bucharest
University for helpful discussions during the course of this work.

\newpage

\begin{center}
\bf{Figure captions}
\end{center}
\bigskip
\medskip

Figure 1:
Energy dependence of the absolute value of the concentration of primary 
defects per unit fluence induced by pions in silicon.

\medskip

Figure 2:
Concentration of vacancies created by irradiation versus time, normalised to 
the concentration of vacancies for the first stage, for different particle fluences, expressed in cm-2.

\medskip

Figure 3:       
Characteristic time of the first stage versus the irradiation fluence, 
for -20, 0 and +20 $^o$C temperature.

\medskip

Figure 4:
a) Time evolution of the concentration of vacancies, normalised to the 
concentration of vacancies remained after the first stage when divacancy 
formation and decomposition are considered, for different particle 
fluences, expressed in cm$^{-2}$.
b) Same as  (a) for the concentration of divacancies.

\medskip

Figure 5: 
a) Time evolution of the concentration of vacancies, normalised to the 
concentration of vacancies remained after the first stage when $V-P$ 
complex formation and decomposition are considered. The three curves 
correspond to different initial impurity concentrations, expressed in 
atomic fractions.
b) Same as (a) for the concentration of complexes.

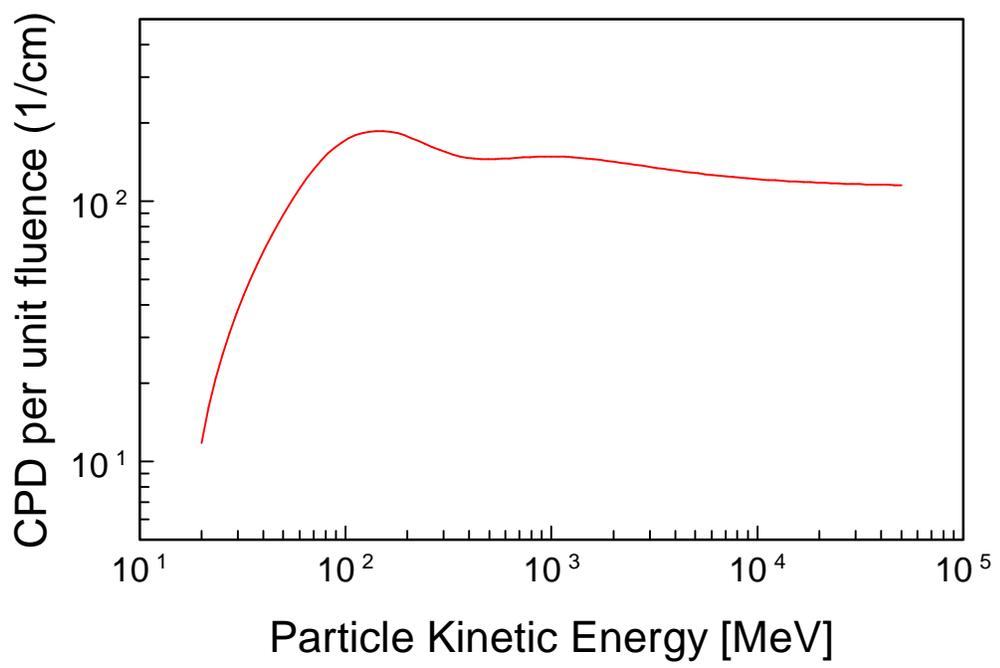
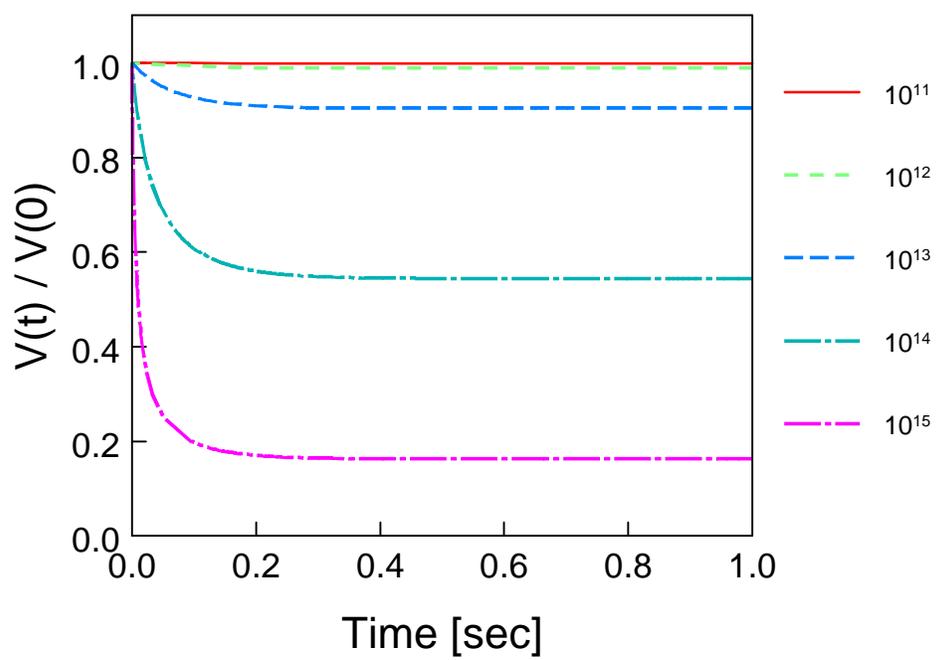
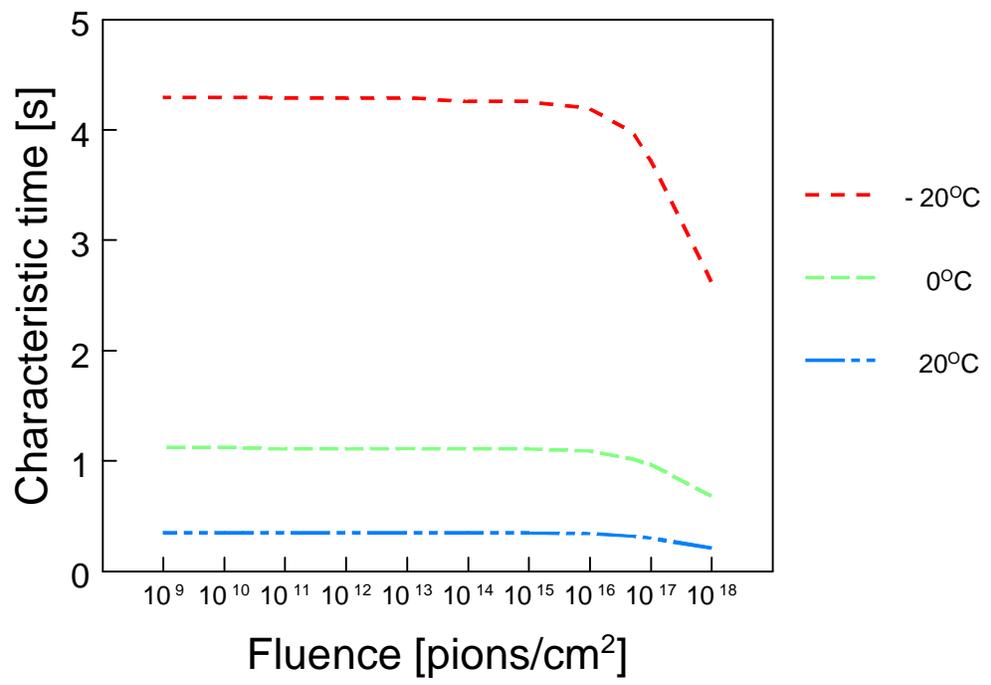
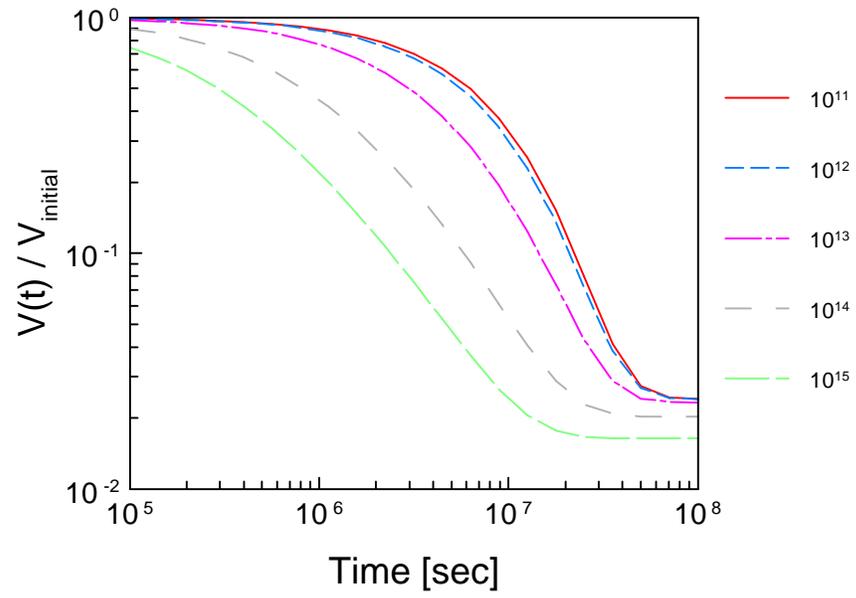
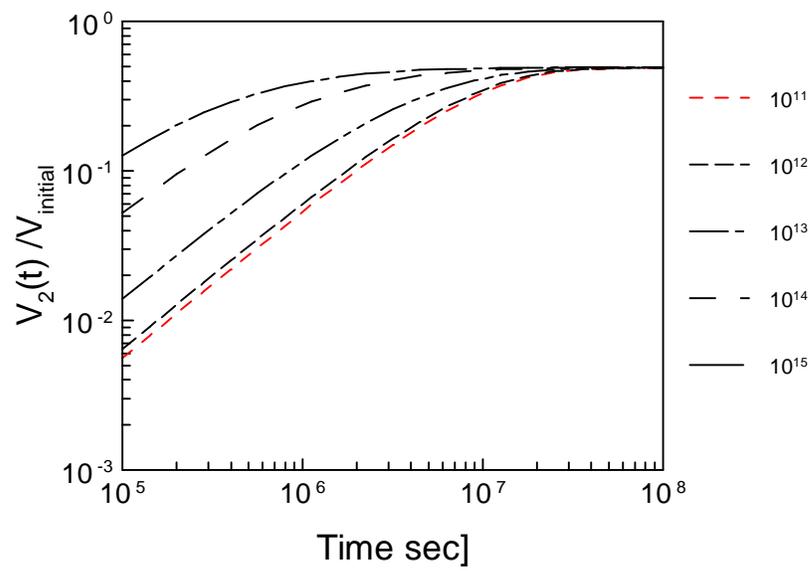
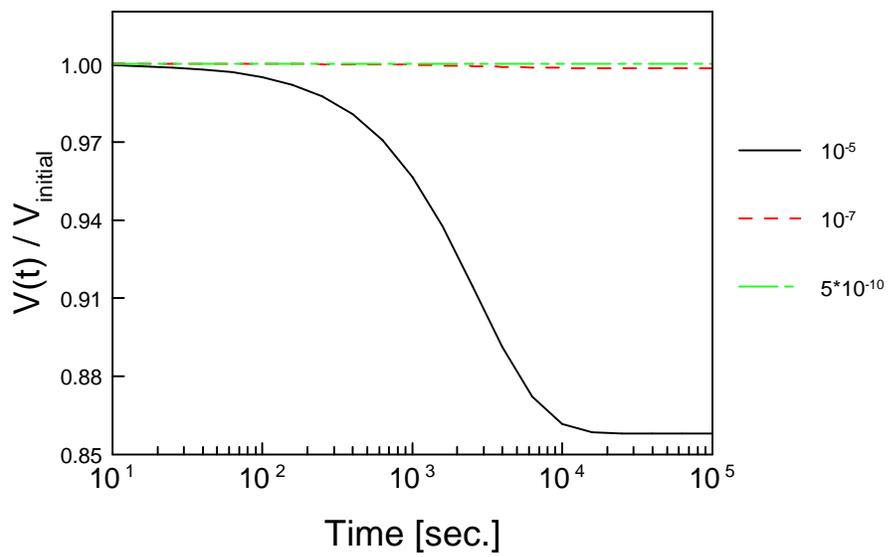
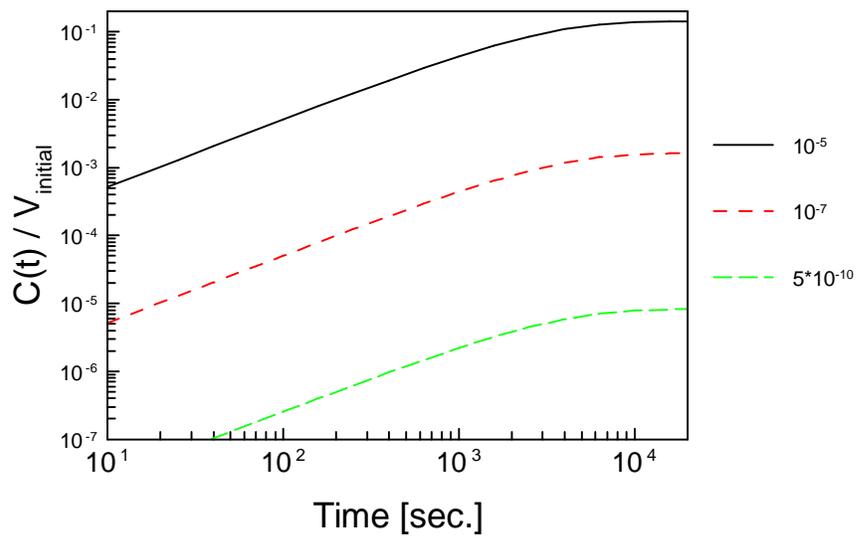
\end{document}